\input harvmac
\input amssym

\def\unit{\relax{\rm 1\kern-.26em I}}
\def\nada{\relax{\rm 0\kern-.30em l}}
\def\tilde{\widetilde}



\def\det{{\rm det}}

\noblackbox
\def\IL{\relax{\rm I\kern-.18em L}}
\def\IH{\relax{\rm I\kern-.18em H}}
\def\IR{\relax{\rm I\kern-.18em R}}
\def\IC{\relax\hbox{$\inbar\kern-.3em{\rm C}$}}
\def\IZ{\relax\ifmmode\mathchoice
{\hbox{\cmss Z\kern-.4em Z}}{\hbox{\cmss Z\kern-.4em Z}} {\lower.9pt\hbox{\cmsss Z\kern-.4em Z}}
{\lower1.2pt\hbox{\cmsss Z\kern-.4em Z}}\else{\cmss Z\kern-.4em Z}\fi}
\def\CM {{\cal M}}

\def\CF {{\cal F}}

\def\CM {{\cal M}}

\def\det{{\rm det}}
\def\Tr{{\rm Tr}}

\font\manual=manfnt \def\dbend{\lower3.5pt\hbox{\manual\char127}}

\def\IZ{\relax\ifmmode\mathchoice
{\hbox{\cmss Z\kern-.4em Z}}{\hbox{\cmss Z\kern-.4em Z}} {\lower.9pt\hbox{\cmsss Z\kern-.4em Z}}
{\lower1.2pt\hbox{\cmsss Z\kern-.4em Z}}\else{\cmss Z\kern-.4em Z}\fi}

\def\lfm#1{\medskip\noindent\item{#1}}

\def\rt2{\sqrt{2}}
\def\irt2{{1\over\sqrt{2}}}

\def\slashchar#1{\setbox0=\hbox{$#1$}           
   \dimen0=\wd0                                 
   \setbox1=\hbox{/} \dimen1=\wd1               
   \ifdim\dimen0>\dimen1                        
      \rlap{\hbox to \dimen0{\hfil/\hfil}}      
      #1                                        
   \else                                        
      \rlap{\hbox to \dimen1{\hfil$#1$\hfil}}   
      /                                         
   \fi}

\def\foursqr#1#2{{\vcenter{\vbox{
    \hrule height.#2pt
    \hbox{\vrule width.#2pt height#1pt \kern#1pt
    \vrule width.#2pt}
    \hrule height.#2pt
    \hrule height.#2pt
    \hbox{\vrule width.#2pt height#1pt \kern#1pt
    \vrule width.#2pt}
    \hrule height.#2pt
        \hrule height.#2pt
    \hbox{\vrule width.#2pt height#1pt \kern#1pt
    \vrule width.#2pt}
    \hrule height.#2pt
        \hrule height.#2pt
    \hbox{\vrule width.#2pt height#1pt \kern#1pt
    \vrule width.#2pt}
    \hrule height.#2pt}}}}
\def\psqr#1#2{{\vcenter{\vbox{\hrule height.#2pt
    \hbox{\vrule width.#2pt height#1pt \kern#1pt
    \vrule width.#2pt}
    \hrule height.#2pt \hrule height.#2pt
    \hbox{\vrule width.#2pt height#1pt \kern#1pt
    \vrule width.#2pt}
    \hrule height.#2pt}}}}
\def\sqr#1#2{{\vcenter{\vbox{\hrule height.#2pt
    \hbox{\vrule width.#2pt height#1pt \kern#1pt
    \vrule width.#2pt}
    \hrule height.#2pt}}}}

\def\figin{\epsfcheck\figin}\def\figins{\epsfcheck\figins}
\def\epsfcheck{\ifx\epsfbox\UnDeFiNeD
\message{(NO epsf.tex, FIGURES WILL BE IGNORED)}
\gdef\figin##1{\vskip2in}\gdef\figins##1{\hskip.5in}
\else\message{(FIGURES WILL BE INCLUDED)}%
\gdef\figin##1{##1}\gdef\figins##1{##1}\fi}
\def\DefWarn#1{}
\def\figinsert{\goodbreak\midinsert}
\def\ifig#1#2#3{\DefWarn#1\xdef#1{fig.~\the\figno}
\writedef{#1\leftbracket fig.\noexpand~\the\figno}%
\figinsert\figin{\centerline{#3}}\medskip\centerline{\vbox{\baselineskip12pt \advance\hsize by
-1truein\noindent\footnotefont{\bf Fig.~\the\figno:\ } \it#2}}
\bigskip\endinsert\global\advance\figno by1}

\lref\RayWK{
  S.~Ray,
  ``Some properties of meta-stable supersymmetry-breaking vacua in Wess-Zumino
  models,''
  Phys.\ Lett.\  B {\bf 642}, 137 (2006)
  [arXiv:hep-th/0607172].
}

\lref\DineGM{
  M.~Dine, J.~L.~Feng and E.~Silverstein,
  ``Retrofitting O'Raifeartaigh models with dynamical scales,''
  Phys.\ Rev.\  D {\bf 74}, 095012 (2006)
  [arXiv:hep-th/0608159].
}

\lref\ShihAV{
  D.~Shih,
  ``Spontaneous R-symmetry breaking in O'Raifeartaigh models,''
  JHEP {\bf 0802}, 091 (2008)
  [arXiv:hep-th/0703196].
}

\lref\IntriligatorPY{
  K.~A.~Intriligator, N.~Seiberg and D.~Shih,
 ``Supersymmetry Breaking, R-Symmetry Breaking and Metastable Vacua,''
  JHEP {\bf 0707}, 017 (2007)
  [arXiv:hep-th/0703281].
}

\lref\FerrettiEC{
  L.~Ferretti,
 ``R-symmetry breaking, runaway directions and global symmetries in
 O'Raifeartaigh models,''
  JHEP {\bf 0712}, 064 (2007)
  [arXiv:0705.1959 [hep-th]].
}

\lref\RayWQ{
  S.~Ray,
  ``Supersymmetric and R symmetric vacua in Wess-Zumino models,''
  arXiv:0708.2200 [hep-th].
}

\lref\AbeAX{
  H.~Abe, T.~Kobayashi and Y.~Omura,
 ``R-symmetry, supersymmetry breaking and metastable vacua in global and local
 supersymmetric theories,''
  JHEP {\bf 0711}, 044 (2007)
  [arXiv:0708.3148 [hep-th]].
}

\lref\FerrettiRQ{
  L.~Ferretti,
  ``O'Raifeartaigh models with spontaneous R-symmetry breaking,''
  AIP Conf.\ Proc.\  {\bf 957}, 221 (2007)
  [J.\ Phys.\ Conf.\ Ser.\  {\bf 110}, 072011 (2008)]
  [arXiv:0710.2535 [hep-th]].
}

\lref\CheungES{
  C.~Cheung, A.~L.~Fitzpatrick and D.~Shih,
  ``(Extra)Ordinary Gauge Mediation,''
  JHEP {\bf 0807}, 054 (2008)
  [arXiv:0710.3585 [hep-ph]].
}

\lref\AldrovandiSC{
  L.~G.~Aldrovandi and D.~Marques,
  ``Supersymmetry and R-symmetry breaking in models with non-canonical Kahler
  potential,''
  JHEP {\bf 0805}, 022 (2008)
  [arXiv:0803.4163 [hep-th]].
}

\lref\GiveonWP{
  A.~Giveon, A.~Katz and Z.~Komargodski,
 ``On SQCD with massive and massless flavors,''
  JHEP {\bf 0806}, 003 (2008)
  [arXiv:0804.1805 [hep-th]].
}

\lref\CarpenterWI{
  L.~M.~Carpenter, M.~Dine, G.~Festuccia and J.~D.~Mason,
  ``Implementing General Gauge Mediation,''
  arXiv:0805.2944 [hep-ph].
}

\lref\DienesGJ{
  K.~R.~Dienes and B.~Thomas,
  ``Building a Nest at Tree Level: Classical Metastability and Non-Trivial
  Vacuum Structure in Supersymmetric Field Theories,''
  arXiv:0806.3364 [hep-th].
}

\lref\SunNH{
  Z.~Sun,
  ``Continuous degeneracy of non-supersymmetric vacua,''
  arXiv:0807.4000 [hep-th].
}

\lref\IntriligatorFE{
  K.~Intriligator, D.~Shih and M.~Sudano,
  ``Surveying Pseudomoduli: the Good, the Bad and the Incalculable,''
  arXiv:0809.3981 [hep-th].
}

\lref\MarquesVA{
  D.~Marques and F.~A.~Schaposnik,
 ``Explicit R-Symmetry Breaking and Metastable Vacua,''
  JHEP {\bf 0811}, 077 (2008)
  [arXiv:0809.4618 [hep-th]].
}

\lref\SunVA{
  Z.~Sun,
 ``Tree level Spontaneous R-symmetry breaking in O'Raifeartaigh models,''
  JHEP {\bf 0901}, 002 (2009)
  [arXiv:0810.0477 [hep-th]].
}

\lref\AmaritiUZ{
  A.~Amariti and A.~Mariotti,
 ``Two Loop R-Symmetry Breaking,''
  arXiv:0812.3633 [hep-th].
}

\lref\MarquesYU{
  D.~Marques,
 ``Generalized messenger sector for gauge mediation of supersymmetry breaking
 and the soft spectrum,''
  arXiv:0901.1326 [hep-ph].
}

\lref\IntriligatorDD{
  K.~A.~Intriligator, N.~Seiberg and D.~Shih,
  ``Dynamical SUSY breaking in meta-stable vacua,''
  JHEP {\bf 0604}, 021 (2006)
  [arXiv:hep-th/0602239].
}

\lref\IY{
  K.~I.~Izawa and T.~Yanagida,
  ``Dynamical Supersymmetry Breaking in Vector-like Gauge Theories,''
  Prog.\ Theor.\ Phys.\  {\bf 95}, 829 (1996)
  [arXiv:hep-th/9602180].
}

\lref\IntriligatorPU{
  K.~A.~Intriligator and S.~D.~Thomas,
  ``Dynamical Supersymmetry Breaking on Quantum Moduli Spaces,''
  Nucl.\ Phys.\  B {\bf 473}, 121 (1996)
  [arXiv:hep-th/9603158].
}

\lref\ORaifeartaighPR{
  L.~O'Raifeartaigh,
 ``Spontaneous Symmetry Breaking For Chiral Scalar Superfields,''
  Nucl.\ Phys.\  B {\bf 96}, 331 (1975).
}

\lref\NelsonNF{
  A.~E.~Nelson and N.~Seiberg,
  ``R symmetry breaking versus supersymmetry breaking,''
  Nucl.\ Phys.\  B {\bf 416}, 46 (1994)
  [arXiv:hep-ph/9309299].
}

\lref\KitanoXG{
  R.~Kitano, H.~Ooguri and Y.~Ookouchi,
  ``Direct mediation of meta-stable supersymmetry breaking,''
  Phys.\ Rev.\  D {\bf 75}, 045022 (2007)
  [arXiv:hep-ph/0612139].
}

\lref\CsakiWI{
  C.~Csaki, Y.~Shirman and J.~Terning,
  ``A simple model of low-scale direct gauge mediation,''
  JHEP {\bf 0705}, 099 (2007)
  [arXiv:hep-ph/0612241].
}

\lref\EssigKZ{
   R.~Essig, J.~F.~Fortin, K.~Sinha, G.~Torroba and M.~J.~Strassler,
   ``Metastable supersymmetry breaking and multitrace deformations of
SQCD,''
   arXiv:0812.3213 [hep-th].
}

\lref\AbelJX{
   S.~Abel, C.~Durnford, J.~Jaeckel and V.~V.~Khoze,
   ``Dynamical breaking of U(1)(R) and supersymmetry in a metastable
vacuum,''
   Phys.\ Lett.\  B {\bf 661}, 201 (2008)
   [arXiv:0707.2958 [hep-ph]].
}

\lref\IzawaGS{
   K.~I.~Izawa, Y.~Nomura, K.~Tobe and T.~Yanagida,
   ``Direct-transmission models of dynamical supersymmetry breaking,''
   Phys.\ Rev.\  D {\bf 56}, 2886 (1997)
   [arXiv:hep-ph/9705228].
}

\lref\HabaRJ{
  N.~Haba and N.~Maru,
  ``A Simple Model of Direct Gauge Mediation of Metastable Supersymmetry
  Breaking,''
  Phys.\ Rev.\  D {\bf 76}, 115019 (2007)
  [arXiv:0709.2945 [hep-ph]].
}

\lref\DimopoulosZB{
  S.~Dimopoulos and H.~Georgi,
  ``Softly Broken Supersymmetry And SU(5),''
  Nucl.\ Phys.\  B {\bf 193}, 150 (1981).
}

\lref\ColemanJX{
  S.~R.~Coleman and E.~J.~Weinberg,
 ``Radiative Corrections As The Origin Of Spontaneous Symmetry Breaking,''
  Phys.\ Rev.\  D {\bf 7}, 1888 (1973).
}

\lref\WittenKV{
   E.~Witten,
   ``Mass Hierarchies In Supersymmetric Theories,''
   Phys.\ Lett.\  B {\bf 105}, 267 (1981).
}

\lref\DineZA{
  M.~Dine, W.~Fischler and M.~Srednicki,
  ``Supersymmetric Technicolor,''
  Nucl.\ Phys.\  B {\bf 189}, 575 (1981).
}

\lref\DimopoulosAU{
  S.~Dimopoulos and S.~Raby,
  ``Supercolor,''
  Nucl.\ Phys.\  B {\bf 192}, 353 (1981).
}

\lref\DineGU{
  M.~Dine and W.~Fischler,
  ``A Phenomenological Model Of Particle Physics Based On Supersymmetry,''
  Phys.\ Lett.\  B {\bf 110}, 227 (1982).
}

\lref\NappiHM{
  C.~R.~Nappi and B.~A.~Ovrut,
  ``Supersymmetric Extension Of The SU(3) X SU(2) X U(1) Model,''
  Phys.\ Lett.\  B {\bf 113}, 175 (1982).
}

\lref\AlvarezGaumeWY{
  L.~Alvarez-Gaume, M.~Claudson and M.~B.~Wise,
  ``Low-Energy Supersymmetry,''
  Nucl.\ Phys.\  B {\bf 207}, 96 (1982).
}

\lref\DimopoulosGM{
  S.~Dimopoulos and S.~Raby,
  ``Geometric Hierarchy,''
  Nucl.\ Phys.\  B {\bf 219}, 479 (1983).
}

\lref\MartinZB{
  S.~P.~Martin,
  ``Generalized messengers of supersymmetry breaking and the sparticle mass
  spectrum,''
  Phys.\ Rev.\  D {\bf 55}, 3177 (1997)
  [arXiv:hep-ph/9608224].
}

\lref\GiudiceNI{
  G.~F.~Giudice and R.~Rattazzi,
  ``Extracting Supersymmetry-Breaking Effects from Wave-Function
  Renormalization,''
  Nucl.\ Phys.\  B {\bf 511}, 25 (1998)
  [arXiv:hep-ph/9706540].
}

\lref\DineYW{
  M.~Dine and A.~E.~Nelson,
  ``Dynamical supersymmetry breaking at low-energies,''
  Phys.\ Rev.\  D {\bf 48}, 1277 (1993)
  [arXiv:hep-ph/9303230].
}

\lref\DineVC{
  M.~Dine, A.~E.~Nelson and Y.~Shirman,
  ``Low-Energy Dynamical Supersymmetry Breaking Simplified,''
  Phys.\ Rev.\  D {\bf 51}, 1362 (1995)
  [arXiv:hep-ph/9408384].
}

\lref\AffleckXZ{
   I.~Affleck, M.~Dine and N.~Seiberg,
   ``Dynamical Supersymmetry Breaking In Four-Dimensions And Its
   Phenomenological Implications,''
   Nucl.\ Phys.\  B {\bf 256}, 557 (1985).
}

\lref\SeibergQJ{
  N.~Seiberg, T.~Volansky and B.~Wecht,
  ``Semi-direct Gauge Mediation,''
  JHEP {\bf 0811}, 004 (2008)
  [arXiv:0809.4437 [hep-ph]].
}



\newbox\tmpbox\setbox\tmpbox\hbox{\abstractfont }
\Title{\vbox{\baselineskip12pt }} {\vbox{\centerline{Notes on SUSY and R-Symmetry Breaking}\vskip8pt\centerline{in Wess-Zumino Models}
}}
\smallskip
\centerline{Zohar Komargodski and David Shih}
\smallskip
\bigskip
 \centerline{{\it School
of Natural Sciences, Institute for Advanced Study, Princeton, NJ
08540 USA}}
\vskip 1cm

\noindent   We study aspects of  Wess-Zumino models related to SUSY and R-symmetry breaking at tree-level. We present a recipe for constructing a wide class of tree-level SUSY and R-breaking models. We also deduce a general property shared by all tree-level SUSY breaking models that has broad application to model building. In particular, it explains why many models of direct gauge mediation have anomalously light gauginos (even if the R-symmetry is broken spontaneously by an order one amount). This suggests new approaches to dynamical SUSY breaking which can generate large enough gaugino masses.

\bigskip

\Date{February 2009}

\newsec{Introduction}

\subsec{Motivations}
Recently, generalized O'Raifeartaigh models --  weakly-coupled Wess-Zumino models that break SUSY through tree-level $F$-term vevs -- have received renewed attention \refs{\RayWK \DineGM\ShihAV\IntriligatorPY\FerrettiEC\RayWQ\AbeAX\FerrettiRQ\CheungES\AldrovandiSC \GiveonWP\CarpenterWI\DienesGJ\SunNH\IntriligatorFE\MarquesVA\SunVA \AmaritiUZ-\MarquesYU}. This is due to the realization that, despite appearances, generalized O'Raifeartaigh models can serve as the low-energy description of dynamical SUSY breaking (DSB) in strongly-coupled gauge theories, as happens in \IntriligatorDD\ (see also \refs{\IY,\IntriligatorPU}).

A common feature of generalized O'Raifeartaigh models is that they typically come equipped with a $U(1)_R$-symmetry. This is no surprise, since it follows from the general result of Nelson and Seiberg \NelsonNF, which says that all generic SUSY-breaking models must have an R-symmetry. However, it leads to some tension with basic phenomenology, in that the R-symmetry must be broken in the vacuum in order to allow for (Majorana) gaugino masses in the MSSM.
Since the simplest O'Raifeartaigh models (including the original one \ORaifeartaighPR) do not break R-symmetry spontaneously in the vacuum, it is an interesting challenge to come up with generalized O'Raifeartaigh models that do.

By now, many O'Raifeartaigh models with spontaneous R-symmetry breaking have been constructed. All these models have relied on the same basic approach and made use of the following general property: in every O'Raifeartaigh model, there always exist tree-level flat directions emanating from any local SUSY-breaking vacuum. (This result was apparently known to experts in the past, but the first place we are aware of it being clearly stated and proven is \RayWK.) Such tree-level flat directions are known as pseudomoduli, because they generally receive an effective potential from radiative corrections. Since they are typically charged under the R-symmetry, one can achieve spontaneous R-breaking by inducing vevs for the pseudomoduli via the effective potential. Using this approach, it was shown how to construct models that break R-symmetry at one-loop via the Coleman-Weinberg potential in \ShihAV; the necessary condition was that there had to be a field with R-charge $R\ne 0,2$. Other models were also recently constructed \refs{\GiveonWP,\IntriligatorFE,\AmaritiUZ}\ where a pseudomodulus is pushed away from the origin at two or higher loops.

One of our goals in this paper is to investigate an alternative approach to R-symmetry breaking which has not received much attention to date: tree-level R-symmetry breaking. By this we mean models where R-symmetry is broken independently of the details of the effective potential. Clearly, in order for this to happen, R-charged fields which are not pseudomoduli must obtain vevs at tree-level, so that R-symmetry is broken on the entire pseudomoduli space.

Along the way we will derive many useful results on Wess-Zumino models. We will argue that these results are relevant to models of dynamical SUSY breaking, and in particular we explain a curious feature of direct gauge mediation that has so far been noticed in many examples: a tendency for gaugino masses to come out anomalously small, even if there is no symmetry protecting the mass term. Equipped with these understandings, we will suggest some new  approaches for dynamical SUSY breaking which could avoid the problem of too-light gauginos.

Now we turn to a more detailed summary of our main results.

\subsec{Summary and detailed outline}

Our investigations will be mainly in context of renormalizable Wess-Zumino models, i.e.\ theories of chiral superfields $\phi_i$ with canonical K\"ahler potential and a general superpotential\foot{Although we assume renormalizability for simplicity, in fact many of our results also apply to non-renormalizable superpotentials.}
\eqn\WWZgen{
W = f_i\phi_i+{1\over2}m_{ij}\phi_i\phi_j+{1\over6}\lambda_{ijk}\phi_i\phi_j\phi_k~.
}
We will refer to models of this form that break SUSY as generalized O'Raifeartaigh models, or O'R models for short. In section 2 we set the stage by reviewing some general properties of Wess-Zumino models. First we derive a general result: {\it a massless fermion in Wess-Zumino models implies a massless complex boson in the same chiral multiplet, even if SUSY is broken.} Then we show how this lemma leads to a new derivation of the existence of pseudomoduli in O'R models: \eqn\flatintro{
\phi_i=\phi_i^{(0)} + z F_i~,
}
where $z\in {\Bbb C}$ is any complex number and $F_i$ is the $F$-term expectation value of the field $\phi_i$. The virtue of this derivation is that it makes clear the connection with the massless Goldstino.

The existence of this universal flat direction implies that there is a convenient basis in which the flat direction is parameterized by some chiral superfield $X$ (which also contains the Goldstino), and the directions orthogonal to it are parameterized by fields $\varphi_i$. In this basis the superpotential looks like \RayWK
\eqn\WWZcanintro{
W = f X  + {1\over2}(\lambda_{ab}X+m_{ab})\varphi_a\varphi_b+{1\over6}\lambda_{abc}\varphi_a\varphi_b\varphi_c~.
}
with the SUSY-breaking pseudomoduli space occuring at $\varphi=0$ and $X$ arbitrary. We will refer to \WWZcanintro\ as the ``canonical form" of O'R models, since every O'R model can be brought to this form.

In addition, we argue that {\it generic} O'R models must take a certain form:
 \eqn\WWZgeniiintro{
W = X_i f_i(\varphi_a) + g(\varphi_a)~.
}
where the fields $X_i$ have $F$-term expectation values while the $\varphi_a$ fields do not. This form is useful since it can explicitly exhibit all the symmetries of the model. This is in contrast to  the canonical form \WWZcanintro, which can hide the symmetries of the problem, due to the set of shifts and unitary rotations that are necessary to transform a given model into \WWZcanintro.

In section 3, we examine more precisely what it means for a Wess-Zumino model to break R-symmetry at tree-level. An important point that we will emphasize is that tree-level R-symmetry breaking requires genuine {\it tree-level SUSY breaking} -- i.e.\ the pseudomoduli space must be locally stable (i.e.\ tachyon-free) everywhere. Otherwise one has to compute the effective potential in order to decide whether the vacuum even exists or not.

While there are many well-known examples of tree-level SUSY breaking models -- including the original O'Raifeartaigh model \ORaifeartaighPR, the ``rank-condition" model that arises as the low-energy limit of massive, free-magnetic SQCD \IntriligatorDD, and the ITIY model \refs{\IY,\IntriligatorPU} -- tree-level R symmetry breaking is relatively new. The first example of tree-level R-symmetry breaking was recently constructed in \CarpenterWI\ and expanded upon in \SunVA.
One of our goals in this paper is to generalize these constructions and find new examples. We will provide a recipe for generating a large class of such models. The idea behind the recipe is to first find a model of the form \WWZgeniiintro\ with $g=0$, which respects a $U(1)_R\times U(1)$ symmetry and
breaks SUSY and the $U(1)$ symmetry at tree-level. Then extend the model to include additional fields $\tilde\varphi_n$ with some superpotential
\eqn\WWZgenintroii{
\delta W = g(\varphi_a,\tilde\varphi_n)~,
}
which does not affect the tree-level SUSY-breaking and leaves invariant only some nontrivial $U(1)_R'\subset U(1) \times U(1)_R$. Now the spontaneous $U(1)$ breaking in the $g=0$ model becomes spontaneous $U(1)_R'$ breaking, and we have a theory of tree-level R-symmetry breaking. This framework characterizes the models of \refs{\CarpenterWI,\SunVA}, where the $g=0$ model is essentially the original O'Raifeartaigh model, and $g$ consists of various terms cubic in the fields. It also leads to qualitatively new models of tree-level SUSY and R-symmetry breaking. In particular, we exhibit a model in which there are no $R=2$ fields in the superpotential (and therefore no linear terms).

Finally, in section 4 we deduce a general property of tree-level SUSY-breaking models: the supersymmetric mass matrix of the fluctuations around the pseudomoduli space always has constant determinant, independent of one's location on the pseudomoduli space. In the canonical form \WWZcanintro, these fluctuations are described by the fields $\varphi_a$, and their mass matrix by $\lambda X+m$. Thus, we show that
\eqn\consdeterintro{
\det(\lambda X +  m) = \det\, m~.
}
As we will explain, this property has an immediate application to gauge mediation model building. In many calculable models of direct gauge mediation (see e.g.\ \refs{\KitanoXG\CsakiWI\AbelJX\HabaRJ-\EssigKZ} for recent examples, as well as the much earlier model of \IzawaGS), the hidden sector turns out to break SUSY at tree-level.  Since the formula for the gaugino mass at leading order in SUSY breaking is \eqn\gauginomassintro{m_{\tilde g}\sim f^\dagger{\partial\over \partial X}\log \det(\lambda X+m)~,}
we see immediately from \consdeterintro\ that in all these models gaugino masses vanish at the leading order in SUSY breaking. This curious fact was noticed in the context of specific models, and here we see it is a general property of tree-level SUSY-breaking.

Since tree-level SUSY breaking is necessary for tree-level R-symmetry breaking, this also implies that models of tree-level R-symmetry breaking cannot be useful for direct gauge mediation. (That is, in spite of the fact R-symmetry is spontaneously broken, gaugino masses do not arise at the leading order.) Of course, they can still be useful as SUSY-breaking hidden sectors in a modular model of gauge mediation with a separate messenger sector.

A promising way of generating large enough gaugino masses is to consider theories of DSB where the SUSY-breaking vacuum is not the ground state even in the low-energy renormalizable approximation. Indeed, then there is no reason to expect the pseudomoduli space to be tachyon free everywhere, and then there would be no obstruction to obtaining satisfactorily large gaugino masses. This strategy is different from the conventional approach to DSB. For example, in the meta-stable DSB model of \IntriligatorDD, the renormalizable ``rank-condition" model breaks SUSY at tree-level, and SUSY is only restored ``dynamically" by non-perturbative effects.  In light of these observations, there is no compelling reason to discard models that do not break SUSY in the low energy approximation, and following such guidelines might lead to qualitatively new models of meta-stable dynamical SUSY breaking.

\newsec{Basics of SUSY Breaking in Wess-Zumino Models}

\subsec{Setup, and a lemma}

In this section we will discuss some general features of generalized O'R models.  While this will consist mostly of a review of known facts, some new insights will also emerge in the course of our discussion.

The starting point of our investigation is a general Wess-Zumino model with chiral superfields $\phi_i$ having a canonical K\"ahler potential and a renormalizable superpotential
\eqn\WWZgen{
W = f_i\phi_i+{1\over2}m_{ij}\phi_i\phi_j+{1\over6}\lambda_{ijk}\phi_i\phi_j\phi_k~,
}
The tree-level scalar potential is (subscripts on $W$ indicate derivatives)
\eqn\VWZgen{
V =\sum_i |W_i|^2~.
}
We suppose that $V$ has a SUSY-breaking local minimum at some $\phi_i=\phi_i^{(0)}$.  This imposes a number of constraints on the theory:

\lfm{1.} Obviously, SUSY-breaking requires at least one $W_i\ne 0$. (All derivatives of $W$ are evaluated at $\phi_i^{(0)}$ unless otherwise noted.)

\lfm{2.} $V$ has extremum provided that
\eqn\reqi{
W_{ij}W_j^*=0~,
}
Given that $(\CM_F)_{ij}\equiv W_{ij}$ is also the fermion mass matrix,  \reqi\ is nothing but the tree-level manifestation of the well-known fact that when SUSY is broken, there is a massless Goldstino, and it corresponds to the direction selected out by the nonzero $F$-terms.

\lfm{3.} The quadratic fluctuations around the extremum are described by the  tree-level boson mass-squared matrix,
\eqn\VWZquadgen{
\CM_B^2 = \pmatrix{ \CM_F^* \CM_F & \CF^* \cr \CF & \CM_F\CM_F^*}
}
where $\CM_F$ was defined above, and
\eqn\Fdefs{
\CF_{ij} \equiv W_k^*W_{ijk}
}
is the effect of SUSY-breaking. In a consistent vacuum, $\CM_B^2$ must be positive semi-definite (i.e.\ there cannot be any tachyons).

\bigskip

{}From these constraints, we will deduce a number of interesting properties of generalized O'R models. We start by proving the following lemma, which will be useful imminently as well as later in the paper:
{\it in any SUSY-breaking vacuum of a generalized O'R model, if there is a massless fermion at tree-level, then its scalar superpartner must also be massless at tree-level.}

To prove the lemma, we use the fact that if $M$ is a positive semi-definite hermitian matrix, then $w^\dagger M w =0$ if and only if $Mw=0$. (This fact is trivial to prove once one recalls that $M$ can always be written as $A^\dagger A$ for some matrix $A$.) Now suppose that $\CM_F$ has a zero eigenvector $v$, and consider the norm of the vector $(v,\,v^*)$ w.r.t.\ $\CM_B^2$:
\eqn\vnorm{
\pmatrix{v \cr  v^*}^\dagger  \pmatrix{ \CM_F^* \CM_F & \CF^* \cr \CF & \CM_F\CM_F^*} \pmatrix{v \cr  v^*} = v^T \CF v +c.c.
}
For positive semi-definite $\CM_B^2$, this must vanish, since otherwise we could make it negative by rotating the phase of $v$.
Hence $(v,\,v^*)$ must be a null eigenvector of $\CM_B^2$. Therefore there is a massless boson and this completes the proof of the lemma.\foot{We note that the methods used here are reminiscent of those used by \DimopoulosZB\ to establish the well-known (but different) result that there must always be a scalar lighter than the up quark, if SUSY is broken and mediated at tree-level. We thank N.~Arkani-Hamed for bringing this to our attention.}

Note that one immediate corollary of the lemma is that
\eqn\corollaryi{
\CF v = 0
}
i.e.\ $\CM_F$ and $\CF$ must both have the same null eigenvector. Another comment is that, from the proof of the lemma, it is clear that the assumption of renormalizable superpotential was unnecessary. So the lemma applies equally well to general polynomial $W$.

\subsec{Existence of pseudomoduli spaces}

Since the Goldstino is always massless, its superpartner must also always be massless in the vacuum of any SUSY-breaking WZ model, according to the above lemma. Applying the corollary \corollaryi\ to $v=W_i^*$ leads to our next general condition on the superpotential:
\eqn\reqii{
W_{ijk} W_i^*W_j^*=0~.
}
In fact, for a renormalizable superpotential, the superpartner of the Goldstino is not only massless -- it can be extended to an entire pseudoflat direction. Using \reqi,\reqii, one immediately finds that
\eqn\pmsgen{
\phi_i=\phi_i^{(0)} + z W_i^*~,
}
leaves the tree-level potential unchanged for any $z\in {\Bbb C}$. The existence of a pseudomoduli space in any SUSY-breaking renormalizable WZ model is a central result in the study of these models. Our derivation of this result is similar to the original presentation in \RayWK, although the use of the lemma above is new.

Note that in \pmsgen, the $F$-terms $W_i$ are evaluated at the stationary point $\phi_i^{(0)}$, even though $z$ is arbitrary. In fact, it does not matter where along the pseudomoduli space the $F$-terms are evaluated -- under a shift in $z$, the $F$-terms do not change:
\eqn\Ftermsdelta{
\delta W_i = W_{ij}(zW_j^*) + {1\over2}W_{ijk}(zW_j^*)(z W_k^*) = 0~,
}
where we have used \reqi,\reqii. Thus the $F$-terms are constant along the pseudomoduli direction \pmsgen.\foot{In the appendix, we argue for a much stronger result: the $F$-terms are constant along the {\it entire} pseudomoduli space, even if it contains other directions in addition to \pmsgen.} It follows then that we can perform a unitary transformation independent of our location on the pseudomoduli space, such that only one field $X$ has an $F$-term vev. Shifting the other fields (which we denote by $\varphi_a$) around their vevs leads to
what we will refer to as the {\it canonical form} of SUSY-breaking WZ models:\foot{Under slightly more refined conditions on the scalar potential, a similar canonical form also exists in the case of a non-renormalizable superpotential \RayWK. The same comment holds for the generic form of SUSY breaking (2.12), which we discuss in the next subsection.}
\eqn\WWZcan{
W = X (f+{1\over2}\lambda_{ab}\varphi_a\varphi_b) + {1\over2}m_{ab}\varphi_a\varphi_b+{1\over6}\lambda_{abc}\varphi_a\varphi_b\varphi_c~.
}
In this basis, \pmsgen\ corresponds to $\varphi=0$ and $X\in {\Bbb C}$ arbitrary.
We would like to stress that the set of transformations that bring the superpotential into the canonical form -- and hence the canonical form itself -- need not respect the symmetries of the problem, if there are any.

\subsec{Generic form of generalized O'R models}

Finally, let us use the preceding discussion to deduce one more elementary result in the theory of generalized O'R models. The idea is that for a {\it generic model}, i.e.\ absent any unnatural cancellations, the requirements \reqi,\reqii\ imply that there cannot be any term in the superpotential coupling together two fields both of which get $F$-terms. Given a certain SUSY breaking classical solution, let us split up the fields into $X_i$ which have nonzero $F$-term vevs and $\varphi_a$ which do not. Then every $X_i$ must appear linearly, i.e.
\eqn\WWZgenii{
W = X_i f_i(\varphi_a) + g(\varphi_a)~.
}
We will refer to this as the {\it generic form} of the generalized O'R model. This form is useful, since it can exhibit explicitly all the symmetries of the problem, in contrast with the canonical form \WWZcan. However, not every O'R model must take this form, only every generic model. To illustrate this point, take for instance any model in the canonical form \WWZcan\ and perform a unitary rotation among all the fields $(X,\varphi)$. Then there will typically be many fields with $F$-term vevs, and many couplings between them. However, these couplings will not all be independent -- they will be related to the couplings of the original model via the unitary transformation. So the model will not be generic.

It is interesting that the generic form \WWZgenii\ essentially corresponds to the class of O'R models studied in \IntriligatorPY, with an addition of the term $g(\varphi)$ in the superpotential. In \IntriligatorPY, the $g=0$ models were analyzed as a particularly simple class of O'R models. Here, we see that with the inclusion of $g\ne 0$, they become the most general possible class of (generic) O'R models.

\newsec{Tree-Level SUSY and R-Symmetry Breaking}

\subsec{Definitions}

In the previous section, we reviewed the existence of pseudomoduli spaces in generalized O'R models (and mentioned some other results that emerged in the course of the discussion). To determine the vacuum of the theory, it is necessary to compute and minimize the radiatively-induced effective potential on the pseudomoduli space. At one-loop, this is given by the Coleman-Weinberg potential \ColemanJX\
\eqn\cwpot{V={1\over 64\pi^2}{\rm S}\Tr \CM^4\log{{\CM^2\over \Lambda^2}}~.}
One should keep in mind that the pseudomoduli space need not be a local minimum everywhere -- it could be locally stable in some places and tachyonic in others. The model of \ShihAV\ and many of the EOGM models studied in \CheungES\ have this property. Thus it is important to check that the minimum of the effective potential is situated at a stable place on the pseudomoduli space.

Since the details of the SUSY-breaking vacuum depend on one (or higher) loop corrections, it is valid to wonder what it even means for SUSY or R-symmetry to be broken at tree-level. We will say that a model {\it breaks SUSY at tree-level} if it has a pseudomoduli space that satisfies the following two conditions:

\lfm{1.} The pseudomoduli space is locally stable everywhere.
\lfm{2.}  The radiative potential on the pseudomoduli space rises at infinity everywhere.

\medskip

\noindent In addition, we will say that a model {\it breaks R-symmetry at tree-level} if

\lfm{3.} The pseudomoduli space breaks R-symmetry everywhere.

\medskip

When these conditions are satisfied, it is not necessary to compute the effective potential in detail to be guaranteed that SUSY or R-symmetry must be broken in the vacuum. The second condition does require some knowledge of the radiative potential, but only at large fields, where it can generally be computed using wavefunction renormalization, using e.g.\ the techniques developed in \IntriligatorFE. Moreover, if the only pseudomodulus is the $X$ field, then the second condition is automatically satisfied. The potential for $X$ always occurs at one loop (unless $X$ is a completely decoupled Polonyi field), and it always rises like $\log X$ times a positive coefficient which is essentially the anomalous dimension of $X$ \WittenKV.

Note that the pseudomoduli space in question need not be the global minimum of the potential; the theory could have multiple disconnected pseudomoduli spaces, or SUSY vacua, or runaway directions. All that is necessary for the definition is that the theory have at least one pseudomoduli space satisfying the conditions above.

\subsec{A general recipe for tree-level SUSY and R-symmetry breaking}

In this subsection, we will provide a general recipe for how to construct models of tree-level SUSY and R-symmetry breaking. While the recipe does not necessarily lead to the most general possible model, we believe it encompasses a wide variety of cases. In particular, it includes the models constructed in \refs{\CarpenterWI,\SunVA}, as well as a new class of models that we will describe in the next subsection. (The reader who is only interested in the applications of tree-level SUSY and R-breaking to model building is advised to skip ahead to section 4, which does not depend on the results presented in the remainder of this section.)

Recall the generic form of O'R models \WWZgenii, which we repeat here for convenience:
\eqn\WWZgenagain{
W = X_i f_i(\varphi_a) + g(\varphi_a)~.
}
The scalar potential of this model takes the form
\eqn\VWZgeniii{
V = \sum_i |f_i(\varphi)|^2 + \sum_a |X_i\partial_a f_i(\varphi) + \partial_a g(\varphi)|^2~.
}
Our recipe stems from the following observation: if $g=0$,  then the model cannot break R-symmetry at tree-level. To see this, notice that when $g=0$, the theory automatically has an $R$-symmetry under which $R(X_i)=2$ and $R(\varphi_a)=0$. Moreover, the condition that the $\varphi$ $F$-terms vanish implies that $X_i$ must be a zero eigenvector of the matrix $M_{ai}\equiv \partial_a f_i$. But then rescaling $X_i$ by any amount leaves the vacuum energy unchanged, so the origin $X_i=0$ must be a connected part of any pseudomoduli space. Since the $X_i$ are the only fields that carry R-charge, this implies that the theory can never break R-symmetry at tree-level.\foot{Although it does not directly concern us here, it is amusing to note that $g=0$ models cannot break the R-symmetry at one-loop either, since all the fields have $R=0$ or $R=2$  \ShihAV.} Thus for tree-level $R$-symmetry breaking, $g\ne 0$ is essential.

Now let us formulate our recipe for constructing models of tree-level SUSY and R breaking. Suppose we have a model with $g=0$ that:

\lfm{1.} Breaks SUSY at tree-level. (This is easier than it may sound: any $g=0$ model that breaks SUSY is guaranteed to have a stable pseudomoduli space.)

\lfm{2.} Respects an ordinary $U(1)$ symmetry in addition to the R-symmetry described above.

\lfm{3.} Stabilizes the $\varphi$ fields at a nonzero value, such that the extra $U(1)$ is spontaneously broken.

\medskip

\noindent Then to obtain a model of tree-level R breaking, we can attempt to add fields $\tilde\varphi$ and a function $g(\varphi,\tilde\varphi)$ in the superpotential, such that  the $U(1)_R$ and $U(1)$ symmetries are broken explicitly, but a nontrivial combination of the two (call it $U(1)_R'$) is left intact. Now, as long as the $\tilde\varphi$ $F$-terms can all be set to zero by solving for $\tilde\varphi$, then the analysis of the model proceeds as for $g=0$ (in particular, SUSY is still broken at tree-level). The only difference is that now the R-symmetry is spontaneously broken by the fact that $\varphi\ne 0$ in the vacuum.

Roughly speaking, we can think of this class of tree-level R-symmetry breaking models as ordinary O'R models ($g=0$) dressed up with extra couplings ($g\ne 0$) whose purpose is to fix the R-charges of all the fields to ``exotic" values.

To illustrate this, let us turn the recent work of \refs{\CarpenterWI,\SunVA}. There models of tree-level R-symmetry breaking were constructed which were essentially of the form
\eqn\CDFM{
W = X_0(f+\lambda\varphi_1\varphi_2)+m(X_1\varphi_1+X_2\varphi_2)+ ({\rm cubic\,\, terms})~.
}
Aside from the cubic terms, the superpotential is basically that of the original O'R model (which breaks SUSY at tree-level), augmented with a $U(1)$ symmetry under which $\varphi_1$, $X_2$ have charge $+1$ and $\varphi_2$, $X_1$ have charge $-1$. When $\lambda f > m^2$, the SUSY-breaking pseudomoduli space has $\varphi_i\ne 0$, breaking the $U(1)$ \IntriligatorPY. So all the conditions above are satisfied, and as long as the cubic terms are chosen correctly (with the inclusion of additional fields),  R-symmetry can be broken at tree-level. In this way, we see that models of \refs{\CarpenterWI,\SunVA}\ fit into our general framework.

\subsec{Example of an O'R model with no $R=2$ fields}

To further illustrate the utility of our general recipe, we will use it to construct a tree-level SUSY and R-symmetry breaking model in which there is no field with R-charge $R=2$ (and hence no linear term in the superpotential). Such theories are interesting since if SUSY is broken then R-symmetry is broken as well by the $F$-terms (none of which are neutral) \RayWQ. The model we construct here is the first explicit example of its kind and should be taken as an existence proof.

Consider the following $g=0$ model:
\eqn\WnoRex{
W = m_1 X_1\varphi_1+m_2 X_2\varphi_2 +\lambda_1X_3\varphi_1^2+\lambda_2 X_1\varphi_2^2+\lambda_3 X_4\varphi_3^2
+\lambda_4 X_2\varphi_1\varphi_3~.
}
This is the most general superpotential consistent with an R-symmetry with charges $R(X)=2$ and $R(\varphi)=0$, and an extra $U(1)$ symmetry with charges
\eqn\qcharges{\eqalign{
 & q(X_1)=-2,\quad q(X_2)=-1,\quad q(X_3)=-4,\quad q(X_4)=2\cr
 & q(\varphi_1)=2, \quad q(\varphi_2)=1,\quad q(\varphi_3)=-1\cr
   }}
The tree-level potential for this model is:
 \eqn\VtreenoRex{\eqalign{
 & V = |m_1\varphi_1+\lambda_2\varphi_2^2|^2+|m_2\varphi_2+\lambda_4\varphi_1\varphi_3|^2+|\lambda_1\varphi_1^2|^2+|\lambda_3\varphi_3^2|^2 \cr
  &\qquad + |m_1X_1+2\lambda_1X_3\varphi_1+\lambda_4X_2\varphi_3|^2
   + |m_2X_2+2\lambda_2X_1\varphi_2|^2 + |2\lambda_3X_4\varphi_3+\lambda_4X_2\varphi_1|^2~.
 }}
Since there is no $R=2$ field, there is a SUSY moduli space: $\varphi_i=X_1=X_2=0$ with arbitrary $X_3$ and $X_4$. We are interested if there is also a SUSY-breaking local pseudo-moduli space with $\varphi_i\ne 0$.

To answer this question, clearly it suffices to set the $\varphi$ $F$-terms (the second line of \VtreenoRex) to zero by an appropriate choice of $X_1$, $X_2$ and $X_4$. This leaves a one-dimensional pseudo-moduli space parameterized by arbitrary $X_3$. So we focus our attention on the reduced potential
  \eqn\VtreenoRexred{\eqalign{
 & V = |m_1\varphi_1+\lambda_2\varphi_2^2|^2+|m_2\varphi_2+\lambda_4\varphi_1\varphi_3|^2+|\lambda_1\varphi_1^2|^2+|\lambda_3\varphi_3^2|^2~. \cr
 }}
The reduced potential has an extremum at (for simplicity we will take all the couplings to be real and positive)
\eqn\varphiextremum{\eqalign{
 & \varphi_2 = -{\lambda_4^2x(2x^2+1)\over \lambda_3 m_2}\varphi_1^2\cr
& \varphi_3={\lambda_4x\over\lambda_3}\varphi_1\cr
& \varphi_1 ={ 2m_2^2\over \lambda_2m_1} {x^2(\alpha^2-x^4)\over (2x^2+1)(x^4-2x^6-2\alpha^2) }
}}
with $\alpha = \lambda_1\lambda_3/\lambda_4^2$ and $x$ a real solution of the equation
\eqn\xeq{
\beta (x^4-2x^6-2\alpha^2)^{3/2} = {4(\alpha^2-x^4)^2x^2\over 2x^2+1}~,
}
with $\beta = m_1^2 \lambda_2\lambda_3/\lambda_4^2 m_2^2$.

The equation for $x$ has real solutions provided that $\alpha$ is sufficiently small. For instance, for $\alpha=0.1$ and $\beta=5$, the equation is solved for $x=\pm 0.43$ and $x=\pm 0.59$. In this case, the latter vacua have a tachyonic direction, while the former vacua are local minima -- all the fluctuations around these vacua are non-tachyonic. Thus the former vacua give rise to the stable, tree-level SUSY-breaking pseudomoduli space that we are after.

Since $\varphi\ne 0$ in the pseudomoduli space, the extra $U(1)$ \qcharges\ is broken in the vacuum. The theory does not break R-symmetry at tree-level, since at the origin of pseudomoduli space the R-symmetry is unbroken. But if we now add the right fields $\tilde\varphi$ and the right function $g(\varphi,\tilde\varphi)$, we can turn the $U(1)$ breaking into R breaking, as described in the previous subsection.

There are many possible ways to do this; one way is to introduce new fields $\tilde\varphi_1$, $\tilde\varphi_2$ and add to \WnoRex\ the superpotential
\eqn\deltaWnoRex{
g(\varphi,\tilde\varphi) = m_3\tilde\varphi_1^2+m_4\tilde\varphi_2\varphi_3+\lambda_5\tilde\varphi_1\tilde\varphi_2^2~.
}
These terms break the $U(1)_R$ and $U(1)$ symmetries but leave intact a nontrivial linear combination
\eqn\qchargesii{\eqalign{
 & R(X_1)=5,\quad R(X_2)={7\over2},\quad R(X_3)=8,\quad R(X_4)=-1\cr
 & R(\varphi_1)=-3, \quad R(\varphi_2)=-{3\over2},\quad R(\varphi_3)={3\over2},\quad R(\tilde\varphi_1)=1,\quad R(\tilde\varphi_2)={1\over2}
   }}
It is important that the most general possible superpotential allowed by this R-symmetry is \WnoRex\ plus \deltaWnoRex. The new $F$-terms $F_{\tilde\varphi_1}$ and $F_{\tilde\varphi_2}$ can be set to zero by choice of $\tilde\varphi_1$ and $\tilde\varphi_2$. So the analysis of the previous model carries over, except that now the R-symmetry \qchargesii\ is spontaneously broken at tree-level.

In the pseudomoduli space of this model, there are three massless modes in the tree-level spectrum. Two modes obviously correspond to the pseudomodulus direction, while the third corresponds to the Goldstone boson of the broken R-symmetry. This should be contrasted with models where the pseudomoduli space contains an R-symmetric point. In such models, the R-Goldstone is sometimes a part of the canonical pseudomodulus space \pmsgen, and there are only two massless modes, since the only source of R-breaking comes from the pseudomodulus itself.
This difference between tree-level and non-tree-level R-breaking models is clearly a general feature, and we discuss it in more detail in the appendix.

\newsec{Application to Model Building}

Our interest in tree-level SUSY and R-symmetry breaking models stems from their potential usefulness in supersymmetric model building and phenomenology, especially models of gauge mediation \refs{\DineZA\DimopoulosAU\DineGU\NappiHM\AlvarezGaumeWY-\DimopoulosGM}. In this section, we will deduce a general feature of tree-level SUSY breaking models that has immediate, broad implications for model building. Our result follows from the definition of tree-level SUSY breaking given in subsection 3.1.

It will be useful to work in the canonical basis of fields \WWZcan, which we repeat here for convenience (with slightly modified notation):
\eqn\supgensectwo{W=f X+{1\over2}(\tilde\lambda_{ab}X+\tilde m_{ab})\varphi_a\varphi_b +{1\over6}g_{abc}\varphi_a\varphi_b\varphi_c~.}
We ask, under what conditions is the pseudomoduli space spanned by $X$ locally stable for any $X$?
By a unitary rotation of the fields, we can always bring $\tilde\lambda$ and $\tilde m$ to the form
\eqn\lmred{
\tilde\lambda = \pmatrix{ \lambda & 0\cr 0  & 0},\qquad \tilde m=\pmatrix{ m & 0 \cr 0 & 0}
}
so that the reduced determinant $\det(\lambda X+ m)$ is nonvanishing for generic $X$.  (If $\tilde m$ or $\tilde \lambda$ is non-degenerate, then $\tilde \lambda=\lambda$ and $\tilde m=m$.) We will show that in fact, the reduced determinant must be a constant function of $X$:
\eqn\consdeter{
\det(\lambda X +  m) = \det\, m~.
}
As we will see, if \consdeter\ is not satisfied, then there are tachyons around the values of $X$ where the determinant vanishes. Keep in mind that \consdeter\ is only a necessary condition, since there could be some light states as $X\rightarrow\infty$ and one has to check them too in order to ensure local stability. This can be done, but it will not be important for our purposes here.

To prove \consdeter, we will use the lemma from section 2. Suppose \consdeter\ does not hold; then since the determinant of $\lambda X+m$ must be a polynomial in $X$,
\eqn\deter{\det(\lambda X+m)=\sum c_i(\lambda,m)X^i~,}
there must be places in the complex $X$ plane where it vanishes. Consider the theory around some such point $X=X_0$, and let $v$ satisfy
\eqn\vdefnull{
(\lambda X_0+  m)v=0~.
}
This corresponds to a massless fermion direction. The lemma that we proved in section~2 implies that either the corresponding boson $\delta\varphi_i=v_i$ must also be massless, or there is a tachyon. But the former option implies, via \corollaryi\ and \vdefnull, that $\lambda v =  m v =0$, which contradicts the assumption that $\det(\lambda X+ m)$ is not identically zero. Therefore, there must be a tachyonic direction in field space around $X=X_0$, but this
contradicts  our assumption of local stability.
We conclude that $\det(\lambda X+ m)$ cannot have any zeroes at finite points in fields space; the only possibility is that it is a constant function. This proves the desired result.

The result \consdeter\ has immediate consequences for models of gauge mediation where the hidden sector is described by a generalized O'R model. In such models, some subset of the $\varphi_a$ fields are charged under the SM gauge group $G_{SM}=SU(3)\times SU(2)\times U(1)$ (with $X$ neutral) and  function as messengers communicating SUSY-breaking to the MSSM. The mass matrix of messengers must factorize at the quadratic order from the other fields due to gauge invariance, so if the hidden sector breaks SUSY at tree-level, this results in
\eqn\consdetermess{\det (\lambda X+m)\bigr|_{messengers}=const~.}
However, this immediately implies that the gaugino masses in the MSSM vanish to leading order in the SUSY-breaking, since at this order they are given by \refs{\CheungES, \GiudiceNI}:
\eqn\gauginomass{m_{\tilde g}\sim f^\dagger{\partial\over \partial X}\log \det(\lambda X+m)\bigr|_{messengers} = 0~.}
Since there is no such cancellation for the sfermion masses, this generally implies that the gauginos are much lighter than the sfermions in such models.\foot{One might hope that for large SUSY-breaking parameters (i.e.\ very low scale gauge mediation) the hierarchy might disappear. However, it generally turns out that even when SUSY breaking is not small, the spectrum has gauginos much lighter than scalars, due to the fact that the higher-order corrections to the gaugino masses are usually not significant in such theories of messengers (see e.g.\  \MartinZB).} This simple result shows that it is impossible to build viable theories of gauge mediation with tree-level SUSY breaking, unless one is prepared to accept an exacerbated little hierarchy problem and the attendant fine tuning coming from very heavy sfermions.

Examples of models with generalized O'R hidden sectors and nonzero gaugino masses include classic constructions such as minimal gauge mediation \refs{\DineYW,\DineVC}. Here the messenger fields are ``put in by hand" and couple to the SUSY-breaking $X$ field only via a cubic superpotential term $\delta W = \lambda X\tilde\varphi^2$, so they always become tachyonic around $X=0$. Other examples include the direct gauge mediation models studied in \CheungES, where the messengers are required for SUSY-breaking, and where there are again always tachyons around $X=0$ if gaugino masses are generated at leading order. Our result elucidates the basic mechanism in these (and other) examples. The gaugino masses can be nonzero at leading order because the pseudomoduli space is not locally stable everywhere.

Although our framework and discussions pertain to classical properties of renormalizable Wess-Zumino models, the main results of this section apply broadly to many models of gauge mediation with dynamical SUSY breaking. Calculable models of DSB, such as \refs{\IntriligatorDD\IY-\IntriligatorPU},  are often described at low energies by a renormalizable WZ model with a large global symmetry group. Gauging a $G_{SM}$ subgroup, one obtains a model of ``direct gauge mediation" \AffleckXZ\ which falls under our framework. It has been observed in many explicit examples \refs{\KitanoXG\CsakiWI\AbelJX\HabaRJ\EssigKZ-\IzawaGS}\ that even if R-symmetry is spontaneously broken, the leading order gaugino masses vanish. The reason for this is obvious from our results: in the models in question, the starting point was always the lowest-lying space of classical vacua. As such, our theorem implies that the model cannot generate leading-order gaugino masses at any point on it, regardless of how the R-symmetry is broken.\foot{In fact, in \KitanoXG\ the authors were led to consider a state with higher energy by explicitly calculating the leading contribution to gaugino masses and observing that it vanishes.} Thus, we see how a phenomenological problem common to many direct mediation models can be understood as a general symptom of tree-level SUSY breaking.

It is intriguing that, in this somewhat limited context, we have arrived at another argument for the inevitability of metastability, which complements the more general argument presented in \IntriligatorPY.
Moreover, our line of reasoning suggests the need for a new type of metastability, not just due to non-perturbative effects as in \IntriligatorDD, but also from states with lower energy within the perturbative, renormalizable approximation.

Obviously, another way to avoid the obstruction to leading-order gaugino masses is to consider hidden sectors which are not described at low-energies by generalized O'R models, e.g.\ strongly-coupled models or models where non-renormalizable K\"ahler or superpotential corrections are important. However, such models are often not calculable, limiting their usefulness for making phenomenological predictions. Also, it is intriguing that when the model is calculable, the gaugino masses often vanish at leading order even though the K\"ahler potential is non-canonical (see e.g.\ the recent example of \SeibergQJ). Perhaps there is a way to generalize our result to the case of non-canonical K\"ahler potential; the fact that  the leading order contribution to gaugino masses is a superpotential term in the effective action might be useful for this.

It would be interesting to explore these directions further. Following the avenue outlined by these observations, one may hope to find a new class of theories that break SUSY dynamically (in a metastable state) and successfully generate viable soft masses.

 \bigskip

\noindent {\bf Acknowledgments:}

We are grateful to N.~Arkani-Hamed and N.~Seiberg for useful discussions and comments on the manuscript. The work of ZK and DS was supported in part by NSF grant PHY-0503584. Any opinions, findings, and conclusions or recommendations expressed in this
material are those of the author(s) and do not necessarily reflect
the views of the National Science Foundation.

\appendix{A}{Collected Results on O'Raifeartaigh Models}

\subsec{Massless particles}

As was mentioned in section 3.3, generalized O'R models which break R-symmetry at tree-level possess a qualitative feature distinguishing them from other models in which R-symmetry is broken by radiative corrections or theories with explicit breaking of R-symmetry. It turns out that if the pseudomoduli space breaks R-symmetry there are necessarily at least three real massless modes -- two from the canonical pseudomodulus \pmsgen\ and one from the Goldstone boson of spontaneous R-symmetry breaking. This is to be contrasted with theories where the origin is part of the pseudomoduli space (like the original O'Raifeartaigh model); in these theories, the R-Goldstone mode is in general identical to the phase of the complex pseudomodulus, so there are only two massless degrees of freedom.

To prove our claim, we need to show that if an O'R model of the form \WWZcan\ breaks R-symmetry at tree-level in some vacuum $\phi_i\ne 0$, then the canonical pseudomodulus and the Goldstone boson of R-symmetry breaking
\eqn\Raction{\phi_i\rightarrow e^{i\theta R_i}\phi_i = \phi_i+i\theta R_i\phi_i+{\cal O}(\theta^2)
}
are independent massless degrees of freedom. Note that this is a local statement about the theory around the point $\phi_i$; what we need is that to linear order around the vacuum, the canonical pseudomodulus and the R-Goldstone boson are linearly independent.

We will prove this claim by contradiction -- if the pseudomodulus and the R-Goldstone boson are linearly dependent somewhere on the pseudomoduli space, then the pseudomoduli space must contain the R-symmetric origin. Suppose that the two modes are linearly dependent; then there must exist a $z\in{\Bbb C}^*$ such that
\eqn\lindep{W_i^*=z R_i \phi_i~.}
The condition for an extremum \reqi\ implies that
\eqn\extri{
 R_i \phi_i W_{ij} = 0
 }
which we recognize as the change in $W_j$ under a $U(1)_R$ rotation:
\eqn\rotR{
{d\over d\theta}W_j(e^{iR_i\theta}\phi_i)\Big|_{\theta=0}
 =  i R_i \phi_i W_{ij} = 0~.
 }
Therefore, all the $F$-terms are invariant under rotations by the R-symmetry. Since $W_i$ has definite R-charge $2-R_i$, the only nonzero $F$-terms must be those of fields with R-charge $R=2$. And according to \lindep, this means that the $F$-terms and vevs must be proportional to one another with the same constant of proportionality. We conclude that one can always reach the origin along the canonical pseudomoduli space, which is the desired result.

\subsec{Constant $F$-terms on pseudomoduli spaces}

In this subsection, we will provide a heuristic argument for a claim made in section 2.2: that $F$-term vevs are always constant on the entire pseudomoduli space. This includes not only the canonical pseudomodulus \pmsgen, but also any other pseudomoduli that may be present in the theory.

Suppose there is some pseudomodulus direction $\phi(x)$, with $x$ a real parameter. Proving that the $F$-terms are constant along this direction amounts to proving that
\eqn\Fconst{
{dW_i\over dx} = W_{ij}{d\phi_j\over dx} =0
}
for every $x$.

By definition, there cannot be any tadpoles along the pseudomoduli direction. This amounts to
\eqn\Vreqii{
  {\partial V\over \partial\phi_i} = W_{ij}W_j^* =0
}
for every $x$. It follows then that
\eqn\Vreqiii{
 {d\over dx}\left({\partial V\over\partial \phi_i}\right) = W_{ij}W_{jk}^*{d\phi_k^*\over dx}+W_{ijk}W_j^*{d\phi_k\over dx} =0
}
We recognize here the action of boson mass-squared matrix \VWZquadgen,
\eqn\Vreqiv{
  \pmatrix{ \CM_F^*\CM_F & \CF^* \cr \CF &\CM_F\CM_F^*}\pmatrix{ {d\phi_i/dx}\cr d\phi_i^*/dx} = 0
 }
Indeed, this is equivalent to saying that $\CM_B^2$ must have a zero eigenvector {\it along the entire pseudomoduli space $x$}.
Now how can this happen? It can arise in three ways:

\lfm{1.} By tuning parameters in the theory. We will ignore this case and only focus on generic theories.

\lfm{2.} By spontaneously breaking a global symmetry. Then the massless mode in question is actually a Goldstone boson, not a genuine pseudomodulus.  Here one can have a cancellation between the two terms in \Vreqiii, in which case the $F$-terms are not necessarily constant, $W_{jk}{d\phi_k\over dx}\ne 0$. However, it's clear how they change as we change $x$ -- they rotate by the action of the spontaneously broken symmetry. (One example of this is the $SO(N)$ O'Raifeartaigh model, $W=h X \vec v\cdot\vec v +m \vec v\cdot \vec w + f X$ in the symmetry breaking phase $|hf|>|m^2|$. There $\vec F_{w}\ne 0$ and rotates under the action of the spontaneously broken $SO(N)$ subject to $|\vec F_w\cdot \vec F_{w}| = const$.)

\lfm{3.} By imposing symmetries on the model that force various couplings to vanish, in such a way that $\CM_B^2$ is degenerate for any choice of the remaining, symmetry-respecting couplings. This can only happen if the vector ${d\phi_i\over dx}$ appearing in \Vreqiv\ is annihilated by both $\CM_F^2$ and $\CF$, i.e.\ if the two terms of \Vreqiii\ vanish individually. This immediately implies the desired result \Fconst.

\listrefs
\end